**A. Fenyvesi***

Institute for Nuclear Research, Hungarian Academy of Sciences (MTA Atomki)

Debrecen, Hungary

# Estimations and integral measurements for the spectral yield of neutrons from thick beryllium target bombarded with 16 MeV protons

## HIGHLIGHTS

Thick target spectral yield of p+Be neutrons produced by 16 MeV protons

Neutron spectrum validation via foil activation and irradiation of transistors

Hardness parameter for displacement damage in bulk silicon

## Abstract

Spectral yield of p+Be neutrons emitted by thick (stopping) beryllium target bombarded by 16 MeV protons was estimated via extrapolation of literature data. The spectrum was validated via multi-foil activation method and irradiation of 2N2222 transistors. The hardness parameter (NIEL scaling factor) for displacement damage in bulk silicon was calculated and measured and $\kappa = 1.26 \pm 0.1$ was obtained.

* Corresponding author. Tel.: +36 52 509273

E-mail address: fenyvesi.andras@atomki.mta.hu

## 1. Introduction

A beam line with a neutron source [1] is available in the Cyclotron Laboratory of MTA Atomki (Debrecen, Hungary) for irradiations with high intensity fast neutrons. Neutron production is done by bombarding a thick (stopping) beryllium target with protons or deuterons accelerated by the MGC-20E cyclotron. The intensity of the produced broad spectrum p+Be or d+Be neutrons depends on the energy of the bombarding particles and the current of their beam. At the MGC-20E cyclotron the highest neutron flux can be achieved by producing p+Be neutrons using extracted beams of protons with $E_p$ = 18 MeV energy.

The energy of the beam extracted from the cyclotron is determined by the radio frequency (RF) and the voltage of the dees of the cyclotron. Before 2013 the active components of the final stage RF driver amplifier of the MGC-20E cyclotron were electronic tubes. However, the age degradation of the electronic tubes limited the maximum frequency of the dee voltage power supply and the maximum available energy of the extracted proton beam was only $E_p$ = 16.4 MeV by 2013. Therefore, a new solid state driver RF amplifier had to be developed for the final stage of the RF system. The new driver RF amplifier was installed in 2014. The development was a part of a long lasting program for renewal of the original electronics of the cyclotron and the beam transport system [2, 3].

During the development several irradiation testing of electronics and photonics components and devices were carried out using $E_p$ = 16.4 MeV energy proton beams and a 3 mm thick target beryllium target. The beryllium target was separated from the vacuum system of the MGC-20E cyclotron by a vacuum window (22 μm thick stainless steel foil made of Duratherm-600 alloy). A stream of He-gas cooled the window foil and the bombarded surface of the Be-target. Because of energy loss in the window foil and the He-gas, the energy of the protons that reached the beryllium target was $E_p$ = 16 MeV. It means the tests were carried out using p(16 MeV)+Be neutrons.

It has to be mentioned that the spectrum of p(16 MeV)+Be neutrons can be used for modelling neutron spectra expected at some sites at the future International Experimental Thermonuclear Reactor (ITER) where some control and data acquisition systems will be operated. The expected neutron spectra will cover the $E_n$ = 0 - 16 MeV neutron energy range.

The planning an irradiation testing and evaluation of the results need information on the spectral yield of neutrons used. 15 different neutron spectra have been available in the literature for the $E_p$ = (14.5 – 23) MeV proton energy range. Neither measured nor estimated neutron spectrum was found in the literature for $E_p$ = 16 MeV bombarding energy. Therefore, the estimation was done on the basis of the literature data. Then the estimated spectrum was checked experimentally via irradiating thin activation foils and 2N2222 bipolar transistors.

This paper presents the results of estimations and the integral measurements for the spectral yields of neutrons produced by $E_p$ = 16 MeV protons on thick (stopping) beryllium target.

## 2. Estimation of the neutron spectrum

The measured thick target yield data taken from [4] and extrapolation of the data published in [5] were the alternatives for estimation.

In [4] spectral distributions of thick target yields of p+Be neutrons produced by protons with $E_p$ = 14.8 MeV, 18 MeV and 23 MeV energy were published in graphical form. The data were measured via time-of-flight (TOF) technique.

Also, TOF data were published in [5] for the spectral distribution of the thick target neutron yield of p+Be neutrons at $E_p$ = 17.24, 18.16, 18.45, 19.08, 19.36, 19.92, 20.55, 20.69, 20.97, 21.04 and 22.01 MeV proton energies, respectively. The number of neutron energy groups is 45 that covers the $E_n$ = 0 – 20.5 MeV neutron energy range. The deviation of the total thick target yield data published in [4] and [5] reaches about 20% at $0^o$ beam direction.

There is limited information in the two papers on the yield of neutrons for the $E_n < 0.5$ MeV neutron energy range.

Our earlier multi-foil irradiation experiments had shown that the agreement of the measured and calculated saturation activities was better when the data published in [5] were used for the calculations. Therefore, the neutron spectrum expected at $E_p = 16$ MeV proton energy was estimated via extrapolation of the data of Table 3 published in Ref. [5].

The $Y_n(E_p;E_n)$ neutron yield as a function of the $E_p$ proton energy was fitted by power law as

$$Y_n\left(E_p[MeV]; E_n[MeV]\right) = A(E_n[MeV]) * \left(E_p[MeV]\right)^{b(E_n[MeV])} \tag{1}$$

for each $E_n \geq 0.5$ MeV neutron energy. Both $E_p$ and $E_n$ were given in MeV unit. After obtaining the $A(E_n)$ and $b(E_n)$ parameters the

$$Y_n\left(E_p = 16\ MeV; E_n[MeV]\right) = A(E_n[MeV]) * (16\ MeV)^{b(E_n[MeV])} \tag{2}$$

extrapolation was done for each neutron energy.

## 3. Foil activation measurements

High purity foil samples (see Table 1) were irradiated with p+Be neutrons produced by protons impinging on the beryllium target with $E_p = 16$ MeV energy. The beam current was constant within 1%. The diameter of each foil was 1.3cm. The geometry centers of the foils were at the $\vartheta = 0^o$ direction at s = 13.4 ± 0.2 cm measured from the geometry center of the 3mm thick beryllium target.

The activated foils were counted by two HPGe spectrometers. The efficiencies of the detectors were measured using calibrated standard radioisotope sources. Then logarithmic polynomials

$$ln\left(\varepsilon_{det}(E_\gamma)\right) = \sum_{i=0}^{3} a_i * ln^i(E_\gamma) \quad (3)$$

were fitted to the calculated efficiencies. The fitted logarithmic polynomials were used for interpolating the detection efficiencies for the gamma energies of the gamma lines counted.

The measured gamma spectra were evaluated by the FGM spectrum evaluation software [6]. The net peak areas obtained for the full energy photo-peaks were used for calculating the $A^*_{sat}$ saturation activity per target atom as

$$A_{sat}^* = \frac{\lambda * C(E_\gamma)}{\frac{m}{M} * N_A \varepsilon_{chem.} \varepsilon_{isot.} \varepsilon_b(E_\gamma) \varepsilon_{det}(E_\gamma)(1-e^{-\lambda * t_i}) e^{-\lambda * t_c}(1-e^{-\lambda * t_m})} \quad (4)$$

where

$\lambda = \ln 2/T_{1/2}$ - decay constant of the radioisotope produced in the neutron induced activation reaction and emitting the γ-photons with energy of $E_\gamma$; $T_{1/2}$ is the half life of the radioisotope,

$E_\gamma$ - energy of the gamma photons to be calculated,

C - net peak area of the full energy photo-peak generated by γ-photons with energy of $E_\gamma$,

m - mass of the activated foil,

M - molar weight of the chemical element,

$N_A$ - Avogadro-number,

$\varepsilon_{chem.}$ - chemical abundance of the element in the foil,

$\varepsilon_{isot.}$ - abundance of the target isotope of the activation reaction,

$\varepsilon_b$ - absolute intensity of the γ-photons with $E_\gamma$ energy to be calculated,

$\varepsilon_{det}$ - detection efficiency for the gamma photons with $E_\gamma$ energy,

$t_i$ - irradiation time,

$t_c$ - cooling time after irradiation until starting the counting,

$t_m$ - time of counting.

No correction factor (H) for the irradiation history was necessary (i. e. H = 1) since the current of the proton beam was constant within 1%.

Nuclear data (half lives, energies of the emitted gamma photons and their absolute intensities) needed for calculation of the saturation activities were taken from the evaluation of [7].

## 4. Estimation of the saturation activities

The expected saturation activities for each measured reaction were calculated from the estimated neutron spectrum and the excitation functions as

$$A_{sat} = \frac{m}{M} N_A I_p t_i \int_0^{E_{n;max}} \Phi(E_n)\,\sigma(E_n)\,dE \qquad (5)$$

where

$I_p$ - current of the proton beam,

$t_i$ - irradiation time,

$\Phi(E_n)$ - flux of neutrons per Coulomb in the ($E_{n-1}$;$E_n$) neutron energy interval,

$\sigma(E_n)$ - cross section of the nuclear reaction averaged for the ($E_{n-1}$, $E_n$) neutron energy interval and assigned.

As Table 1 shows the evaluated cross section data were taken from the ENDF/B-VII.1 library [8] or from the IRDFF v.1.05 library [9] for estimating the saturation activities. Linear interpolation was done for calculating the cross sections for the mean energies of each energy group.

## 5. Measurement and calculation of the NIEL scaling factor

Users interested in irradiation testing of electronics frequently need the Non Ionising Energy Loss (NIEL) scaling factor (or hardness factor, κ) for the broad neutron spectrum used for the irradiation testing. The NIEL-scaling concept was developed for bulk silicon first. The $\Phi_{eq.}$(1 MeV) equivalent fluence of 1 MeV neutrons induces the same number of permanent displacements in Silicon than the Φ total fluence of the broad spectrum neutrons used in the tests.

$$\Phi_{eq.}(1MeV) = \kappa\Phi \qquad (6)$$

The $\kappa$ conversion factor can be calculated as

$$\kappa = \frac{1}{D(E_n = 1\ MeV)} * \frac{\int_0^{E_{n;max}} D(E_n)\varphi(E_n)dE_n}{\int_0^{E_{n;max}} \varphi(E_n)dE_n} \qquad (7)$$

where $D(E_n)$ is the displacement damage cross section function for Si and $D(E_n = 1$ MeV$)$ = $9.5\text{x}10^{-2}$ MeV b. Griffin et al. (1993) A compiled recommended damage function for Si has been published in [10]. The tabulated form of the published data was taken from [11].

The neutron spectrum extrapolated for $E_p$ = 16 MeV bombarding proton energy on the basis of the data published in [5] was used for calculating its κ-factor.

The κ-factor has been derived from measurements, too, following the procedure described in the ASTM E1855-05e1 standard test method [12]. 2N2222 NPN bipolar junction transistors (20 pcs) with TO-18 type metal can housing were irradiated. The transistors were used as silicon displacement damage sensors. The small-signal current gain ($h_{FE}$) (the β parameter) was measured before and after the irradiation.

The degradation of $h_{FE}$ is result of the change of bulk concentrations of crystal defects (radiation induced bulk damage) during irradiation. Some types of defects are traps for the charge carriers in the transistor that reduce the lifetime, mobility and concentration of the charge carriers and, thus, $h_{FE}$ of the transistor. The defects are created mainly by recoils from nuclear interactions (elastic and inelastic scattering, nuclear reactions) induced by neutrons. Gamma-photons accompanying neutrons can damage chemical bonds of the silicone crystal and, thus, they can give a contribution, too. Different types of defects with different life times are created.

According to Messenger and Spratt [13] and Messenger and Hubbs [14] the degradation of the inverse $h_{FE}$ is proportional with the equivalent flux of 1 MeV neutrons and the

$$\frac{1}{h_{FE;1}} - \frac{1}{h_{FE;0}} = K * \Phi_{eq.}(1\ MeV) \tag{8}$$

Messenger-Spratt equation is valid, where the quantities are

$h_{FE;0}$ and $h_{FE;1}$ - gain before and after irradiation, respectively,

K - damage constant.

The K calibration factors for the transistors were taken from results of a previous measurement reported in [15]. The calibration factors scattered around K = $4.27*10^{-15}$ $cm^2/n$ with 9.8% standard deviation. The age degradation of the transistors was checked by a control group of transistors. They have never been irradiated and annealed but they were stored together with the measuring transistors always. No age degradation of the transistors was observed.

Before irradiation the transistors were kept at T = 459K (186 $^{o}$C) for t = 24 hours to anneal long lived crystal defects. Then the starting value of $h_{FE;0}$ was measured.

The transistors were arranged in a 6mm high and 3cm dia. cylindrical volume that was fixed on a 1 mm thick metal sheet made of AlMgSi alloy. During irradiation the sheet was perpendicular to the $\vartheta = 0^{o}$ direction and the geometry center of the container was at s = 15.4 cm measured from the geometry center of the 3mm thick beryllium target.

After irradiation the transistors were kept at T = 355K (82 $^{o}$C) for t = 2 hours to anneal unstable short lived defects. Then the degraded value of the current gain was measured ($h_{FE;1}$).

There was no way for calibration of the gamma sensitivities of the transistors to gamma photons. According to the ASTM E1855-05e1 standard test method [12] $K_{\gamma} \cong 1.5*10^{-15}$ $Gy^{-1}$ was supposed. The dose rate of the gamma photons ($2.5*10^{-2}$ $Gys^{-1}$) at the irradiation position was estimated from results of measurements performed using ionization chambers during several former irradiations.

## 6. Results

Figure 1 shows the neutron energy dependence of the parameters used for the power fitting extrapolation of the neutron spectra published in [5].

The spectral distribution of the neutron yield obtained from the extrapolation is shown in Figure 2. The yield integrated for the $E_n$ = 0.4 MeV – 16 MeV neutron energy range is Y = $9.07*10^{15}$ $sr^{-1}C^{-1}$.

Table 2 shows the saturation activities calculated for the different activation reactions using the extrapolated spectrum and the excitation data taken from the evaluated data libraries. The measured saturation activities are also shown in Table 2 with the C/E ratios of the calculated ( C ) and experimental (E) saturation activities. The agreement is within the uncertainty of the calculated and the measured data.

The main relative measurement uncertainties were a) (3-5)% for the detection efficiency of the detector, b) (1–44)% for the counting statistics, c) 4% for the irradiation position of the foils and d) 1% for the measured target charge. The individual relative uncertainties of the parameters used for calculating the measured saturation activities were summed up in quadrature for estimating the relative uncertainties of the measured saturation activities.

In the case of the $^{90}Zr(n,2n)^{89}Zr$ reaction there is a large discrepancy between the calculated and measured saturation activities. The main reasons for the discrepancy are a) the large statistical uncertainty (13%) of the counting statistics, b) the uncertainty of the net peak (29%) area of the $E_\gamma$ = 908.96 keV gamma line of $^{89}Zr$ after background subtraction and c) the large uncertainty of the estimated neutron fluxes above the $E_{th}$ = 9.43 MeV threshold of the reaction.

In the case of the $^{46}Ti(n,p)^{46}Sc$ reaction the main reasons for the large uncertainty of the saturation activity are a) the statistical uncertainty of the counting statistics (6%) and b) the uncertainty (13%) of the net peak area of the $E_\gamma$ = 889.277 keV gamma line of $^{46}Ti$ after background subtraction.

For the hardness parameter (NIEL scaling factor) $\kappa_{calc.}$ = 1.27 was obtained using the extrapolated spectrum and the damage function recommended by Griffin et al. [10].

During the $t_i$ = 11070 sec long irradiation of the transistors Q = 0.205 C total charge was delivered to the beryllium target by the $E_p$ = 16 MeV energy protons. For the position of the transistors $\Phi^{p(16\ MeV)+Be} = 1.05*10^{13}$ n/cm$^2$ energy integrated neutron fluence was calculated from the neutron spectrum extrapolated for $E_p$ = 16 MeV proton energy on the basis of the data published in [5]. The estimated gamma dose was $D_\gamma \approx 300$ Gy ± 20%.

After irradiation and pre-reading annealing of the transistors their $h_{FE;1}$ gains were measured and $\Delta(1/h) = (1/h_{FE;1}) - (1/h_{FE;0}) = 5.62*10^{-2}$ ± 9.3% values were obtained. The sensitivities of the transistors scattered around $K = 4.27*10^{-15}$ $cm^2/n$ with 9.8% standard deviation. The obtained equivalent fluences of $E_n$ = 1 MeV neutrons were $\Phi^{p(16\ MeV)+Be}_{eq.}(1\ MeV) = 1.32*10^{13}$ $n/cm^2$ with 5% standard deviation. Thus, $\kappa_{meas.} = 1.25$ ± 5% was obtained for the hardness factor from the measurements with 2N2222 transistors.

## 7. Conclusions

The thick target spectral yield of p(16 MeV)+Be neutrons produced by $E_p$ = 16 MeV protons has been estimated on the basis of literature data. The estimated spectrum was checked via the foil activation method and irradiating 2N2222 bipolar junction transistors. The hardness parameter of the extrapolated spectrum has been calculated and measured. The agreement of the calculated and measured saturation activities is within the uncertainty of the calculations and measurements. Similarly, the agreement of the calculated and measured hardness parameters is within the uncertainty of the calculations and measurements.

The results can be used for Monte Carlo simulation of irradiations and applications such as radiation hardness tests (bulk displacement damage tests) of electronic devices performed with p+Be neutrons produced at low energy proton accelerators.

Detailed investigation is necessary to estimate covariance matrix of the spectrum and to obtain statistically correct uncertainty for the κ-factor.

**Acknowledgements**

This work was supported from the Research, Technology and Innovation Fund of Hungary by the National Development Agency of Hungary in the framework of the New Széchenyi Plan. ID code of the project: URKUT_10-1-2011-0005. Also, this work was supported by the Hungarian Scientific Research Fund (OTKA). The OTKA Contract No. was NN104543.

Figure 1.

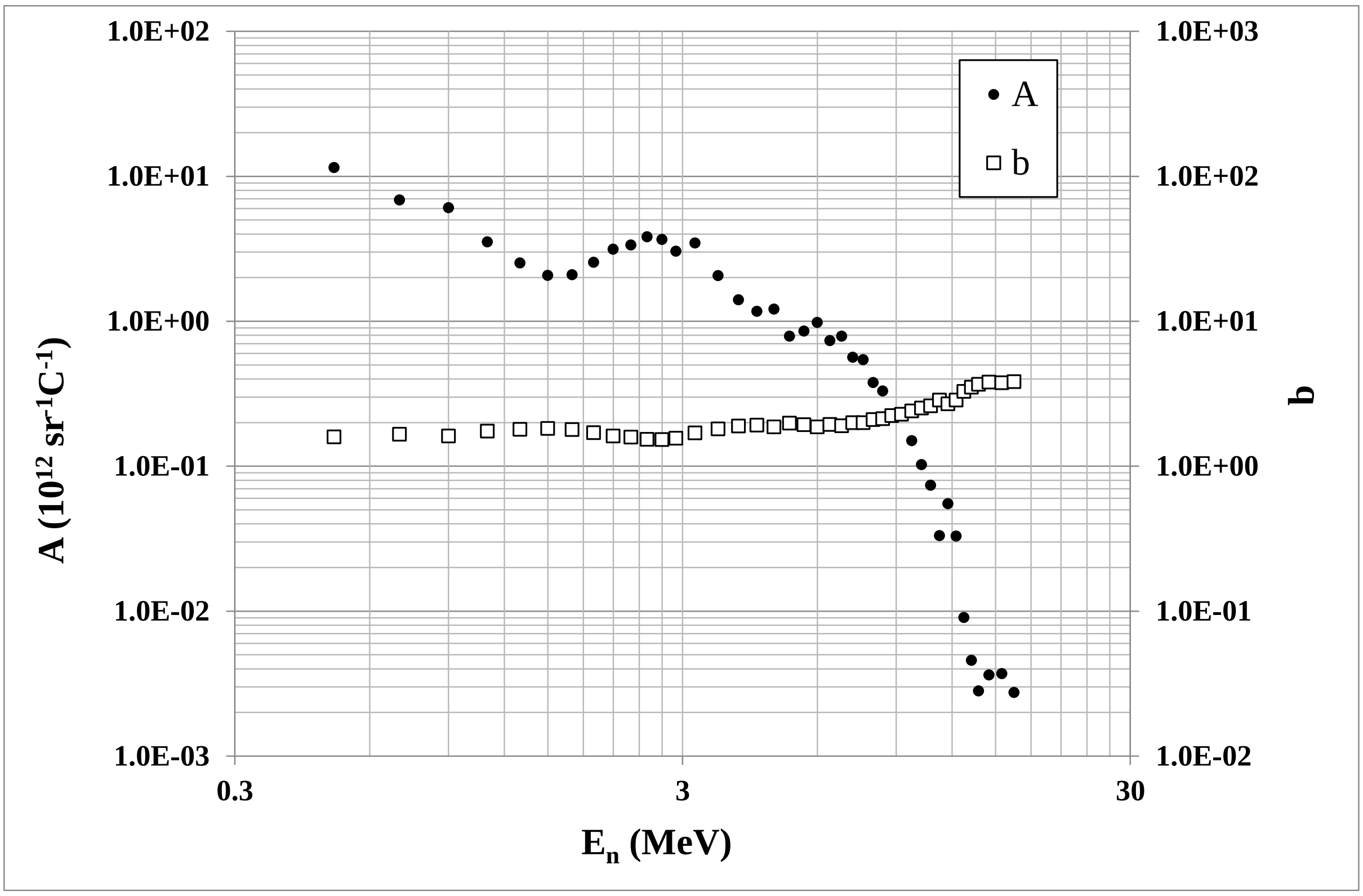

Figure 2

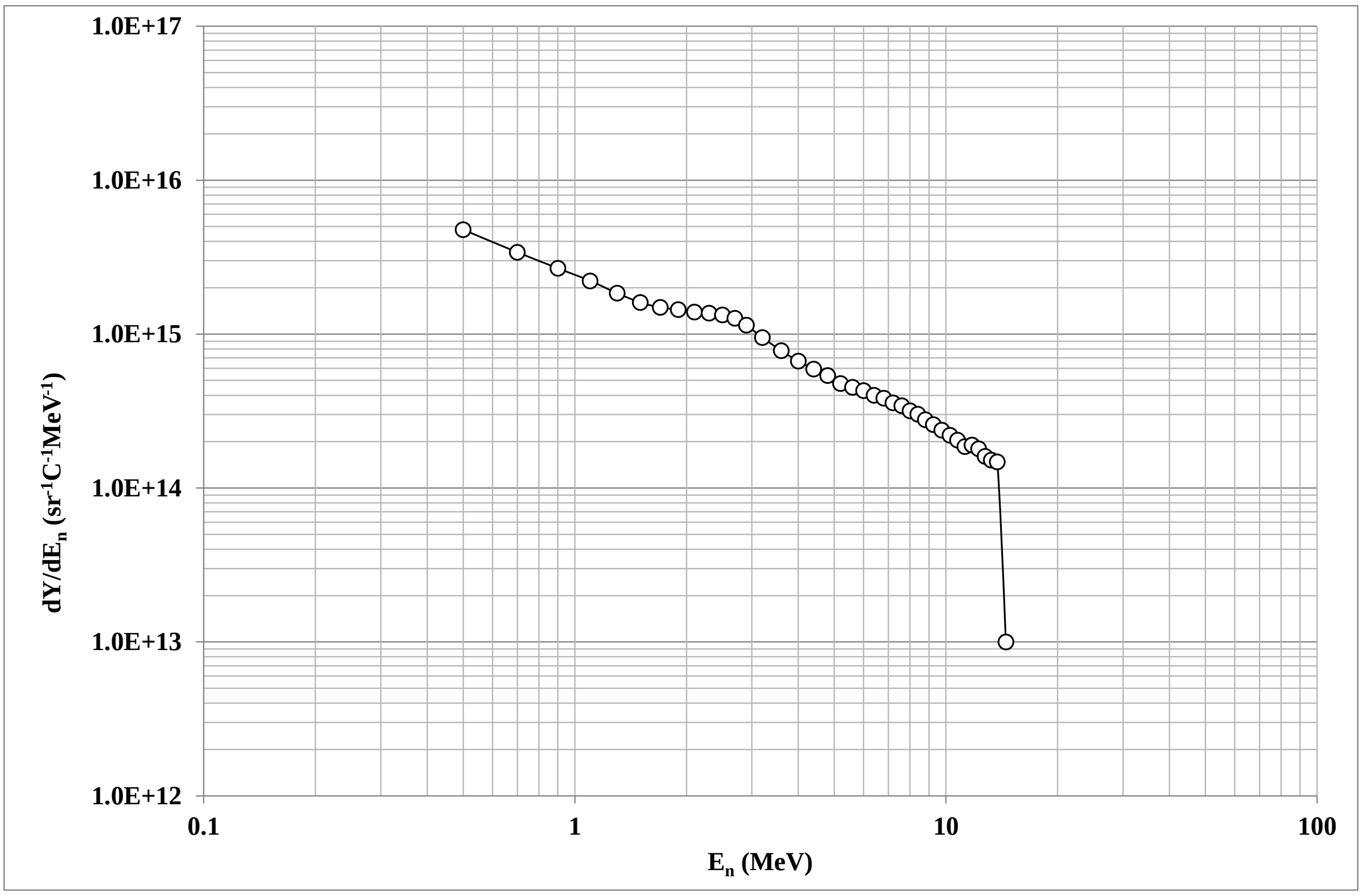

**Table 1.**

| **Foil** | **Diameter** (mm) | **Thicknes** (μm) | **Chemical purity** (%) | **Reactions** | **Evaluated data library** |
|---|---|---|---|---|---|
| Aluminum | 13 | 760 | 99.963 | $^{27}Al(n,p)^{27}Mg$<br>$^{27}Al(n,\alpha)^{24}Na$ | ENDF/B-VII.1<br>ENDF/B-VII.1 |
| Titanium | 13 | 250 | 99.876 | $^{46}Ti(n,p)^{46}Sc$<br><br>$^{47}Ti(n,p)^{47}Sc$<br><br>$^{48}Ti(n,p)^{48}Sc$ | IRDFF v.1.05, 09 October, 2014<br><br>IRDFF v.1.05, 09 October, 2014<br><br>IRDFF v.1.05, 09 October, 2014 |
| Iron | 13 | 130 | 99.984 | $^{56}Fe(n,p)^{56}Mn$ | ENDF/B-VII.1 |
| Nickel | 13 | 250 | 99.981 | $^{58}Ni(n,p)^{58}Co$ | IRDFF v.1.05, 09 October, 2014 |
| Zirconium | 13 | 130 | 99.7392 | $^{90}Zr(n,2n)^{89}Zr$ | ENDF/B-VII.1 |
| Rhodium | 13 | 25 | 99.813 | $^{103}Rh(n,n')^{103m}Rh$ | ENDF/B-VII.1 |
| Indium | 13 | 130 | 99.993 | $^{115}In(n,n')^{115m}In$ | IRDFF v.1.05, 09 October, 2014 |
| Gold | 13 | 25 | 99.91 | $^{197}Au(n,2n)^{196}Au$ | ENDF/B-VII.1 |

**Table 2.**

| | | $A^{*}_{sat}$ (Bq per target atom) | | |
|---|---|---|---|---|
| **Reaction** | **$E_{th}$ (MeV)** | **Calculated** | **Measured** | **C/E** |
| $^{27}$Al(n,p)$^{27}$Mg<br>$^{27}$Al(n,α)$^{24}$Na | 1.90<br>3.25 | 7.35E-18<br>5.43E-18 | 7.50E-18 ± 7%<br>5.24E-18 ± 7% | 0.98<br>1.04 |
| $^{46}$Ti(n,p)$^{46}$Sc | 1.62 | 2.48E-17 | 2.89E-17± 19% | 0.86 |
| $^{47}$Ti(n,p)$^{47}$Sc | 0.00 | 1.81E-17 | 1.85E-17 ± 7% | 0.98 |
| $^{48}$Ti(n,p)$^{48}$Sc | 3.27 | 2.28E-18 | 1.98E-18 ± 7% | 1.15 |
| $^{56}$Fe(n,p)$^{56}$Mn | 2.97 | 5.25E-18 | 5.28E-18 ± 7% | 0.99 |
| $^{58}$Ni(n,p)$^{58}$Co | 0.00 | 9.51E-17 | 1.05E-16 ± 7% | 0.90 |
| $^{90}$Zr(n,2n)$^{89}$Zr | 12.10 | 3.52E-18 | 6.51E-18 ± 48% | 0.54 |
| $^{103}$Rh(n,n')$^{103m}$Rh | 0.00 | 6.79E-16 | 6.91E-16 ± 7% | 0.98 |
| $^{115}$In(n,n')$^{115m}$In | 0.00 | 8.53E-17 | 8.39E-17 ± 7% | 1.02 |
| $^{197}$Au(n,2n)$^{196}$Au | 9.11 | 6.01E-17 | 5.83E-17 ± 7% | 1.03 |

## Figure captions

Figure 1. Parameters of the power law fit as a function of the neutron energy.

Figure 2. Spectral yield of p+Be neutrons produced by 16 MeV protons on thick beryllium target. The tabulated data published in [5] were used for the estimation.

## Table captions

Table 1. Data of the activation foils used in the experiment, the activation reactions and the evaluated data libraries for the cross section data.

Table 2. Calculated ( C ) and measured ( E ) saturation activity per target atom and the C/E values obtained for different nuclear reactions induced by p+Be neutrons emitted by a thick (stopping) beryllium target bombarded with $E_p$ = 16 MeV energy protons.